\documentclass[aps,prl,reprint,superscriptaddress,floatfix]{revtex4-1}

\usepackage{graphicx}
\usepackage[usenames]{color}
\usepackage{dcolumn}
\usepackage[normalem]{ulem}

\def\angstrom{\AA}
\def\inverseangstrom{\AA$^{-1}$}
\def\abx{AB$_2$X$_4$}

\def\irte2{IrTe$_2$}
\def\cis{CuIr$_2$S$_4$}

\def\lro{LiRh$_2$O$_4$}

\def\fd3m{Fd$\overline 3$m}
\def\i41amd{I4$_{1}$/amd}
\def\fmmm{Fmmm}

\graphicspath{{./figures/}}

\begin{document}

\title{Local Structural Evidence for Strong Electronic Correlations in \lro\ Spinel}

\author{K. R. Knox}
\affiliation{Department of Condensed Matter Physics and Materials Science, Brookhaven National Laboratory, Upton, New York 11973, USA}

\author{A. M. M. Abeykoon}
\affiliation{Department of Condensed Matter Physics and Materials Science, Brookhaven National Laboratory, Upton, New York 11973, USA}

\author{H. Zheng}
\affiliation{Materials Science Division, Argonne National Laboratory, Argonne, IL  60439, USA}

\author{W.-G. Yin}
\affiliation{Department of Condensed Matter Physics and Materials Science, Brookhaven National Laboratory, Upton, New York 11973, USA}

\author{A. M. Tsvelik}
\affiliation{Department of Condensed Matter Physics and Materials Science, Brookhaven National Laboratory, Upton, New York 11973, USA}

\author{J. F. Mitchell}
\affiliation{Materials Science Division, Argonne National Laboratory, Argonne, IL  60439, USA}

\author{S. J. L. Billinge}
\affiliation{Department of Condensed Matter Physics and Materials Science, Brookhaven National Laboratory, Upton, New York 11973, USA}
\affiliation{Department of Applied Physics and Applied Mathematics, Columbia University, New York, New York 10027, USA}

\author{E. S. Bozin}
\affiliation{Department of Condensed Matter Physics and Materials Science, Brookhaven National Laboratory, Upton, New York 11973, USA}

\date{\today}

\begin{abstract}
The local structure of the spinel \lro\ has been studied using atomic pair distribution function (PDF) analysis of powder x-ray diffraction data.  This measurement is sensitive to the presence of short Rh-Rh bonds that form due to dimerization of Rh$^{4+}$ ions on the pyrochlore sublattice, independent of the existence of long range order.  We show that structural dimers exist in the low-temperature phase, as previously supposed, with a bond shortening of $\Delta r \sim 0.15$~\AA . The dimers persist up to 350~K, well above the insulator-metal transition, with $\Delta r$ decreasing in magnitude on warming.  Such behavior is inconsistent with the Fermi surface nesting-driven Peierls transition model. Instead, we argue that \lro\ should properly be described as a strongly correlated system.

\end{abstract}

\maketitle

The interplay of charge, spin, orbital and lattice degrees of freedom driven by electron correlation is a unifying principle across a wide range of transition metal systems~\cite{milli;n98,dagot;s05,georg;arocmp13}.  Such a picture provides the framework of understanding needed when one-electron physics is inadequate. In this paper we argue, based on structural principles, that the spinel \lro, whose coupled spin, charge, and lattice behaviors have been understood in terms of single particle physics (band Jahn-Teller transition and charge density wave (CDW) formation~\cite{okamo;prl08}), is better described as a strongly correlated system.

\begin{figure}
\includegraphics[width=85mm]{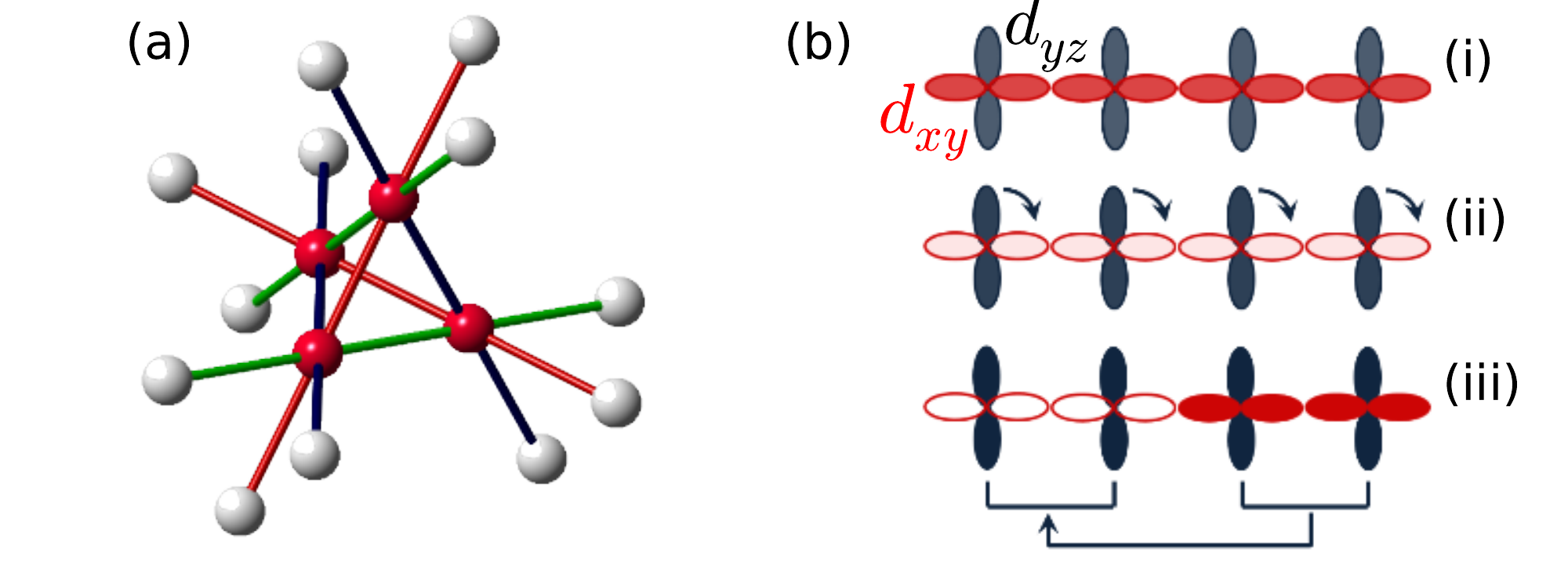}
\caption{\label{fig:model} (Color online) (a) Structural motif of Rh-pyrochlore sublattice. Red atoms denote single Rh$_{4}$ tetrahedron. Red, blue, and green colored bonds differentiate independent chain directions. (b) 2-dimensional sketch of different charge distributions in $d_{xy}$ (red) and $d_{yz}$ (blue) orbitals of the $t_{2g}$ manifold with charge filling represented by color intensity: (i) homogeneous, (ii) sub-band disproportionated (band Jahn-Teller effect), and (iii) CDW. Arrows denote disproportionation of charge.}
\end{figure}

\lro, discovered in 2008~\cite{okamo;prl08}, belongs to a class of spinels described by the common formula \abx. Like \lro\ these systems often exhibit interesting phenomena such as spin frustration, charge, spin and orbital ordering and metal insulator transitions (MIT)~\cite{matsu;prb97,rodri;prl98,ishib;prb02,radae;n02,schmi;prl04,dimat;prl04,takub;prl05,kiryu;prl06,zhou;prl07,erran;jap08}.  For example, spinels with Ti or Ir on the B site undergo a symmetry lowering phase transition from a high temperature metallic phase into a low temperature insulating B-B dimerized state.  This transition has been explained using a band Jahn-Teller model~\cite{okamo;prl08} with dimerization the result of a Peierls transition associated with CDW formation~\cite{khoms;prl05}. The pyrochlore sublattice of corner shared B$_{4}$ tetrahedra, composed of intersecting chains of B-ions as shown in Fig.\ref{fig:model} (a),
%
%
%
%
contains electronically active valence electrons in the $t_{2g}$ orbitals of the $d$-shell.  In the metallic high temperature cubic (HTC) phase these orbitals are triply degenerate~\cite{radae;njp05}, while in the low temperature insulating phase a band Jahn-Teller transition breaks this degeneracy.

In \lro\ the MIT occurs in two stages~\cite{okamo;prl08} and it has been suggested that the two-step nature of the transition could shed light on the mechanism of the transition. At $T_{c1} = 225$~K there is a structural phase transition to an intermediate temperature tetragonal (ITT) phase that breaks the degeneracy of the $t_{2g}$ levels and changes the electron filling of the orbitals oriented along different crystallographic directions.  At $T_{c2} = 170$~K a second transition occurs into a low temperature orthorhombic (LTO) phase. This state is insulating with, presumably, a CDW \emph{along} the chains. In related materials the CDW is seen in average~\cite{radae;n02} and local~\cite{bozin;prl11} structural measurements as a very short metallic B-B bond, although it has not been experimentally established in \lro\ before this work.

The nominal charge of Rh in \lro\ is 3.5+, implying that, on average, there are 5.5 electrons in the $t_{2g}$ manifold since the $e_g$ levels are at higher energy and unpopulated.  If the charge were distributed evenly there would be 0.5 holes per Rh atom shared among the $t_{2g}$ levels (Fig.~\ref{fig:model} (b)(i)).  Charge can disproportionate between the available orbitals in two distinct ways.  First, it can separate between the three $t_{2g}$ orbitals, $d_{xy}$, $d_{yz}$ and $d_{xz}$, on a single atomic site (sub-band disproportionation, Fig.~\ref{fig:model} (b)(ii)).  Second, the charge may distribute itself unevenly between atoms along the chain directions, forming a CDW, regardless of the partitioning of charge between orbitals on an individual atom (Fig.~\ref{fig:model} (b)(iii)). In this paper we will use CDW, dimerization and short-bond interchangeably to describe this since they highlight different aspects of the same phenomenon.  In the extreme case this would result in 3+ and 4+ Rh ions along the chain~\cite{okamo;prl08}.
The LTO structure in \lro\ has not been solved, but in a similar system, \cis , the pattern of charge is nominally -(3+)-(3+)-(4+)-(4+)-~\cite{croft;prb03} with the 4+ ions dimerizing.

The main result of this work is the observation of short Rh-Rh distances over a wide temperature range, which includes all three structural phases.  This means that in \lro\ the dimers are confirmed to exist at low temperature, but also well above the LTO insulating state. Thus, the ITT to LTO transition may be thought of as a crystallization of dimers from a liquid formed at higher temperatures in the metallic phase. We argue that, taken together, all the experimental data are at odds with the existing understanding~\cite{khoms;prl05} where the phase transitions in \abx\ systems are explained by the Peierls mechanism driven by Fermi surface effects.

\lro\ was synthesized using a solid state reaction under high pressure oxygen.  High purity Rh$_2$O$_3$ was mixed with a 10 weight-percent excess of Li$_2$O$_2$.  The mixed powders were fired at 900~$^{\circ}$C for 20 hours under an oxygen pressure of 0.6~MPa.  The resulting powder was found to be single phase by laboratory x-ray diffraction. DC susceptibility data were measured on cooling in a 1~T field using a Quantum Design PPMS.  Resistivity was measured using a standard four-terminal technique.

X-ray total scattering data were collected at the X17A beam line (beam energy 67.42~keV) of the National Synchrotron Light Source (NSLS) at Brookhaven National Laboratory (BNL) in the 80-500~K temperature range using standard protocols~\cite{chupa;jac03, bozin;prl11, egami;b;utbp13}.
Experimental PDFs of \lro\ were obtained by Fourier transforming the reduced total scattering structure function, $F(Q)$, over a broad range of momentum transfer, $Q$ ($Q_{max}  = 28$~\inverseangstrom).

Lattice parameters were obtained using a Le Bail fit to the diffraction data using {\sc GSAS}~\cite{larso;unpub87}, utilizing \fd3m\ (HTC), \i41amd\ (ITT), and \fmmm\ (LTO) models from the literature~\cite{okamo;prl08}. Structural refinement of PDF data was carried out using PDFgui~\cite{farro;jpcm07}.

\begin{figure}
\includegraphics[width=85mm]{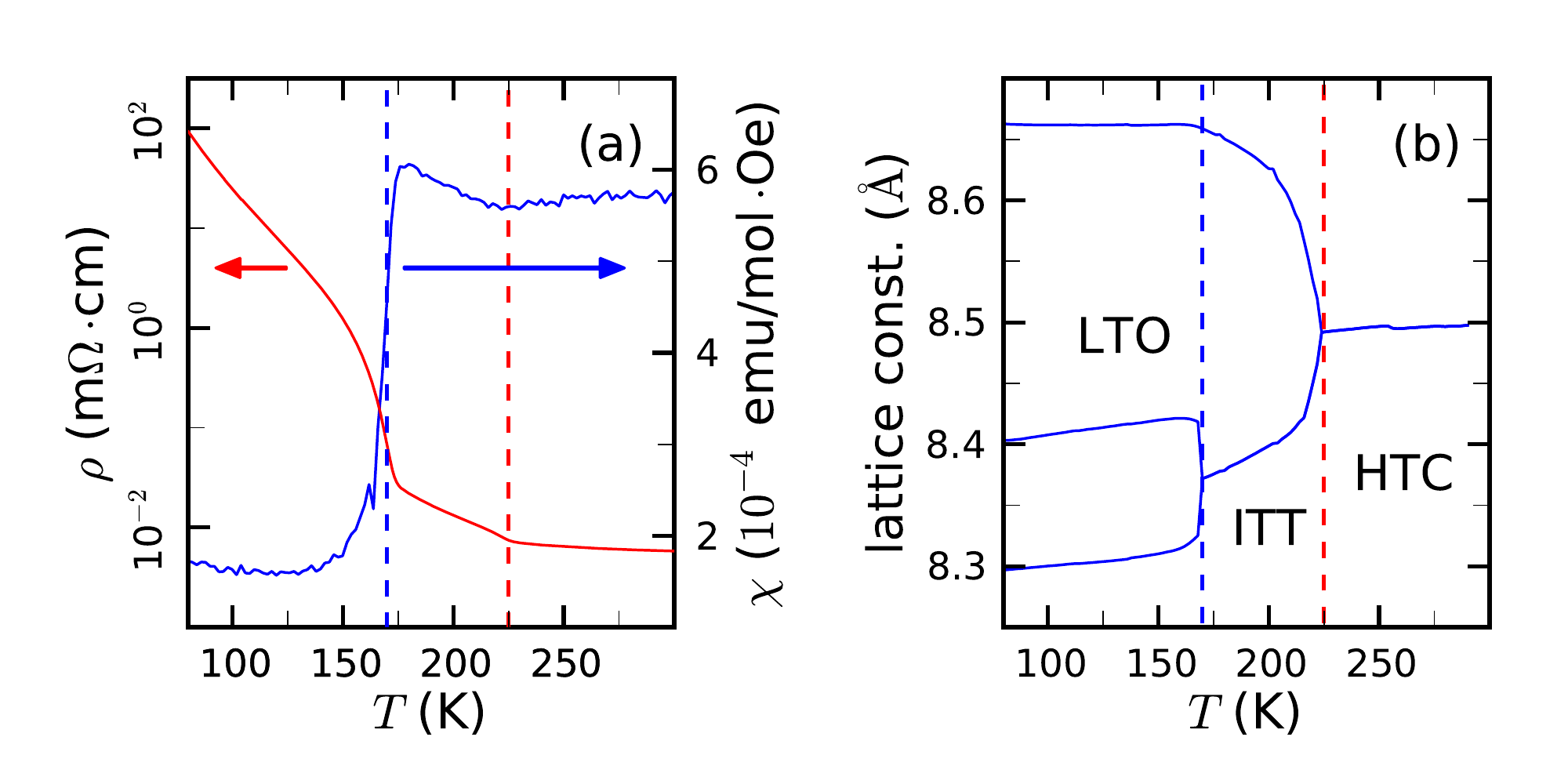}
\caption{\label{fig:LiRhCharacterization} (Color online) (a) Temperature dependence of \lro\ resistivity (red) and magnetic susceptibility (blue). (b) Lattice parameters determined from Le Bail fit of the x-ray data. Vertical dashed lines denote structural phase transitions at 170~K (blue) and 225~K (red).}
\end{figure}

Our sample exhibits resistivity, magnetic susceptibility and average structural changes (shown in Fig.~\ref{fig:LiRhCharacterization}) consistent with earlier reports~\cite{okamo;prl08}.
%
%
The structural phase transitions are clearly observed, with corresponding changes in resistivity and magnetization.

\begin{figure}
\includegraphics[width=85mm]{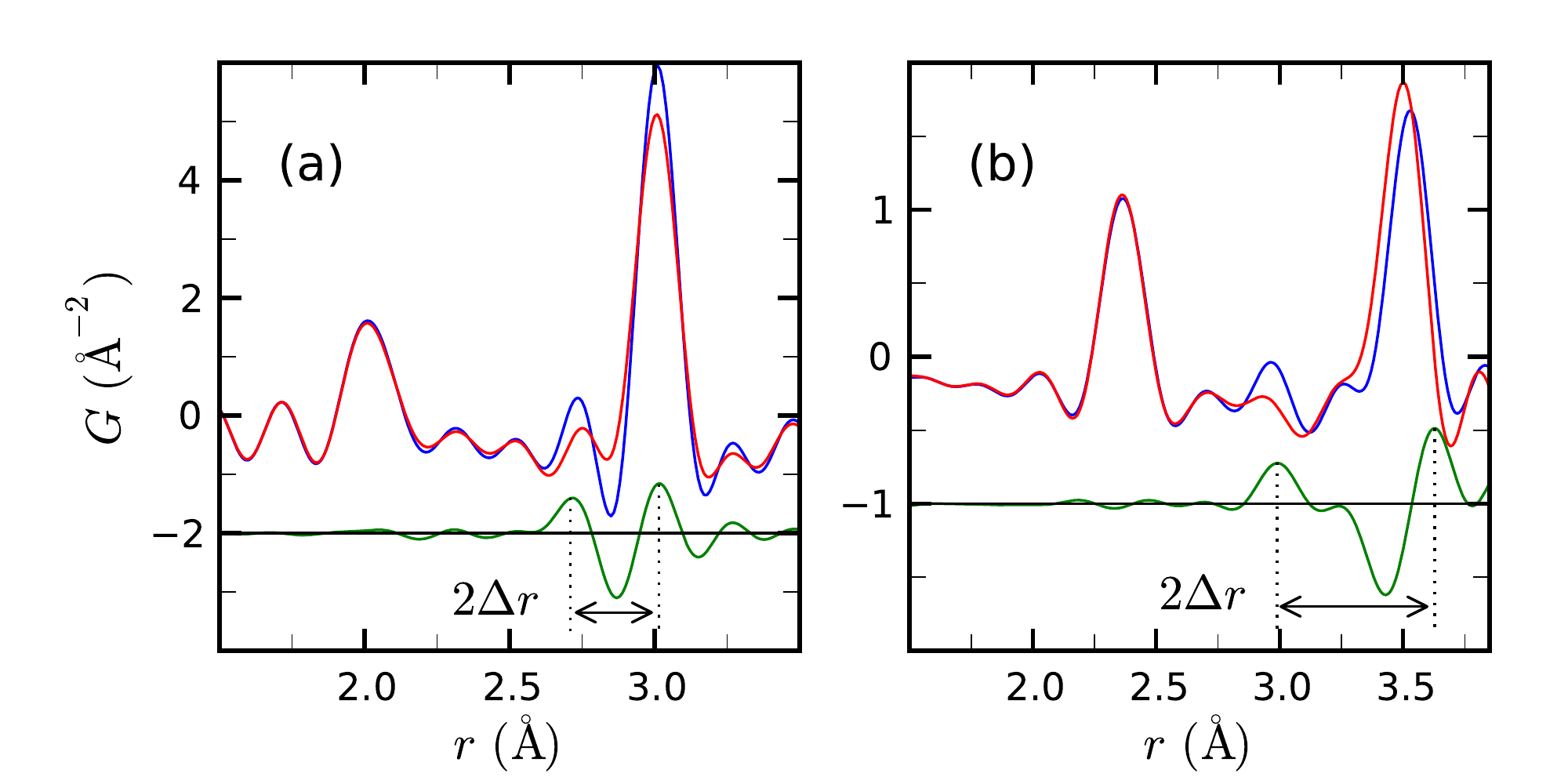}
\caption{\label{fig:dimerAtLT}(Color online) Comparison of experimental PDFs at 240~K (red)
and 160~K (blue) for (a) \lro\ and (b) \cis.  Difference curves (green) are offset for clarity. Dotted lines indicate $r$-position of short (dimerized) and long (undimerized) peaks.  $\Delta r$, defined as half the distance between the long and short bonds, is used as a measure of the distortion associated with dimerization.
}
\end{figure}

Despite the lack of a low temperature crystallographic model, evidence of a dimerized state can be immediately seen by comparing high and low temperature experimental \lro\ PDFs in the low-$r$ regime.
Figure~\ref{fig:dimerAtLT} (a) shows the low-$r$ region of the \lro\ PDF in the LTO and HTC phases along with a difference curve. Since the atomic number of Rh is significantly larger than those of Li and O, the PDF intensity in this region is dominated by Rh-O (2.0 \AA\ peak) and Rh-Rh (3.0 \AA\ peak) atomic pair correlations. In the average HTC model there is only one near-neighbor Rh-Rh distance, corresponding to the sharp PDF peak at 3.0~\angstrom, whereas in the LTO phase the orthorhombic distortion splits the near-neighbor Rh-Rh distance into three average distances, the shortest of which is $\sim$ 2.95~\AA.
However, an additional contribution to the PDF clearly appears at $\sim$ 2.7~\angstrom\ in the LTO phase. A similar observation has been made in \cis\ (Fig.~\ref{fig:dimerAtLT} (b)) where the presence of dimers in the low-temperature phase is well established~\cite{radae;n02,bozin;prl11}.
%
%
%
%
The formation of dimers along 1D chains is characterized by the redistribution of PDF intensity into short (dimerized) and long (non-dimerized) bonds. This occurs with a concomitant loss of intensity at the position of the \emph{average} bond along the 1D dimerizing chain indicating a transfer of intensity from the principal peak to the long and short bond peaks. This redistribution of intensity emerges in Fig.~\ref{fig:dimerAtLT} as a signature M-shape in the difference curve. In \cis\ dimerization occurs along the shorter of two inequivalent Ir chains, which contains the active electronic orbitals that are available for dimerization~\cite{radae;n02}. Similarly, in \lro\ dimerization is expected to occur along the shortest of three Rh chains~\cite{okamo;prl08}. Thus, in both systems the loss of PDF intensity occurs at the leading edge of the Rh-Rh (Ir-Ir) peak at $\sim$3.0~\AA\ ($\sim$3.5 \AA).
The associated structural distortion, $\Delta r$, is $\sim$0.15~\angstrom\ in \lro\, and $\sim$0.45~\angstrom\ in \cis.
This unambiguously establishes the presence of dimers in the LTO phase of \lro, experimentally confirming prior speculation based on anomalies in transport and susceptibility and by analogy with \cis~\cite{okamo;prb00,nakat;prb11}.

\begin{figure}
\includegraphics[width=85mm]{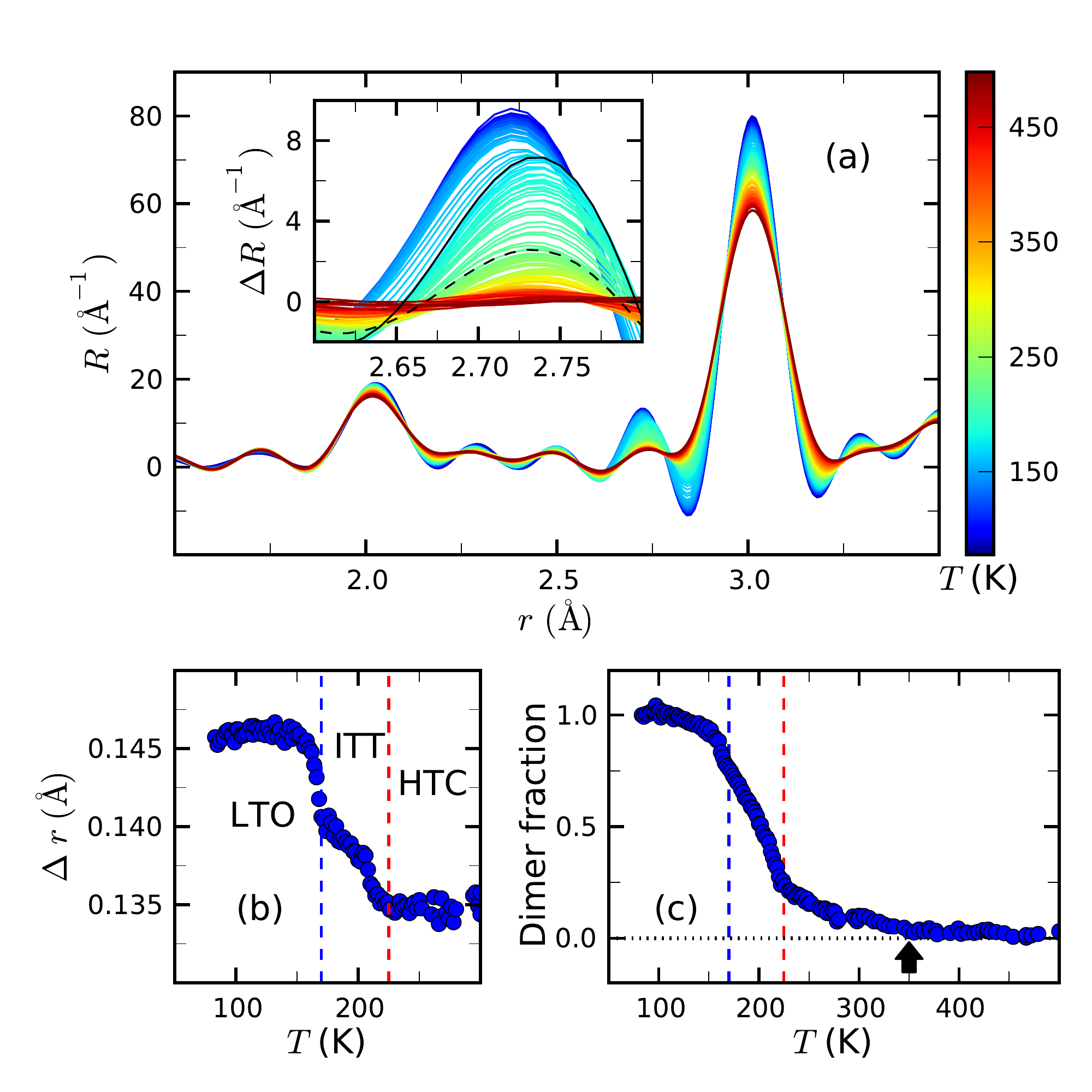}
\caption{\label{fig:RDF} (Color online) (a) Plot of temperature evolution of \lro\ RDF in 1.5 - 3.5~\angstrom\ range. Inset displays differential RDF near dimer peak obtained by subtracting a reference RDF at 500~K with black lines at 170~K (solid), and 225~K (dashed) denoting differential RDFs at phase transition temperatures. (b-c) Results of differential RDF analysis: (b) distortion, $\Delta r$, (c) integrated intensity of short bond peak normalized to its low-T value, representing the dimer fraction.  Black arrow indicates observed break in slope at 350~K. Horizonal dotted line, drawn at zero, is a guide to the eye. Vertical dashed lines indicate phase transition temperatures.
}
\end{figure}

We now turn our attention to the temperature evolution of the Rh-dimers, which is observable in the raw data. Figure~\ref{fig:RDF} (a)
%
%
%
%
shows the temperature evolution of the radial distribution function (RDF), $R(r)$, a related form of $G(r)$~\cite{egami;b;utbp13}; $R(r)\text{d}r$ represents the number of bonds with lengths between $r$ and $r + \text{d}r$.
While the region surrounding the peak at 2~\angstrom\ is largely unchanged, a significant redistribution of PDF intensity occurs in the 2.5 - 3.25~\angstrom\ range with increasing temperature. It is clear from a careful examination of the inset to Fig. \ref{fig:RDF}(a) that this redistribution does not occur abruptly at $T_{c2}$. Rather, the short bond lengthens slightly on warming from 160~K to 170~K. Then, as the sample is heated through the ITT phase and into the HTC phase intensity is gradually transferred from the 2.7~\angstrom\ region to the leading edge of the large peak at 3~\angstrom. This implies that the dimer does not disappear at the LTO to ITT phase transition. Rather, it survives into the ITT and HTC phases with gradually diminishing intensity.

The temperature evolution of the dimers is further quantified by tracking the
peak position and integrated intensity of the relevant $R(r)$ features.
The size of the distortion associated with the dimer formation, $\Delta r$ (see Fig.~\ref{fig:dimerAtLT}), is determined by subtracting a reference RDF at 500~K from all measured RDFs and analyzing the resulting difference curves.
As evident from Fig.~\ref{fig:RDF} (b) $\Delta r$ is nearly constant as the sample is heated through the LTO phase. Then, it changes rapidly in the 160 - 170~K range, decreasing from $\sim 0.146$ to 0.14~\angstrom\ as the LTO/ITT phase transition is approached. As temperature is further increased through the ITT phase $\Delta r$ changes smoothly, gradually decreasing until it stabilizes at 0.133~\angstrom\ at the ITT/HTC phase boundary.

An estimate of the number of dimers present in the sample is obtained by following the integrated intensity of the short-bond peak in the RDF difference curve. As shown in Fig.~\ref{fig:RDF}(c) the dimer fraction is relatively stable through the LTO phase. A rapid drop is observed between 160 and 170~K. Notably, the dimer fraction does not drop to zero at $T_{c2}$, but decreases gradually through the ITT phase. It survives even into the HTC phase, disappearing at $\sim$350~K.

\begin{figure}
\includegraphics[width=85mm]{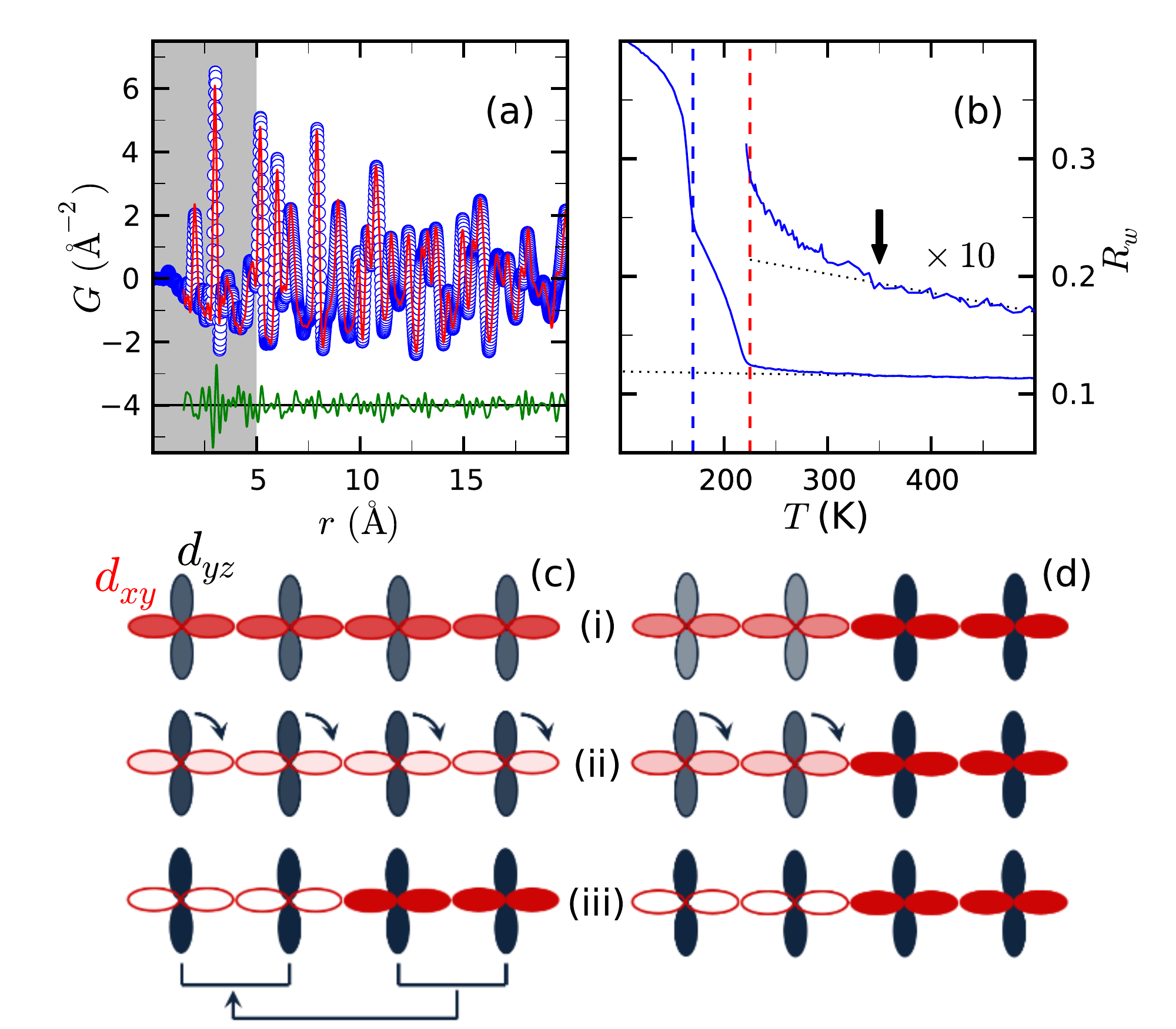}
\caption{\label{fig:PDFfit} (Color online)
(a) Best fit (red line) to \lro\ PDF (blue circles) at 200~K in the 1.5 - 20~\angstrom\ range using \i41amd model.  Difference curve (green) is offset for clarity. Shaded area denotes poorly fit region. (b) Refinement residual, $R_w$, for the entire temperature range. Black dotted line is a linear fit to the high temperature tail. Inset shows expanded view in the 200-500~K range, with observed break in slope at 350~K indicated by arrow. Vertical dashed lines denote phase transitions.
(c) \lro charge distribution model based on band Jahn-Teller scenario: (i) homogeneous charge distribution, (ii) sub-band charge disproportionation and (iii) CDW. (d) Alternative model based on experimental observations made in this work, with CDW existing in all three phases. Like in Fig.~\ref{fig:model} (b), color intensity corresponds to the average amount of charge, while arrows denote charge disproportionation.
}
\end{figure}

Further confirmation of our results is obtained by model dependent analysis of the \lro\ PDFs.
Fig.~\ref{fig:PDFfit} (a) shows a best-fit of the ITT model
to the \lro\ PDF at 200~K in the 1.5 \angstrom\ to 20 \angstrom\ range.
%
%
%
%
The fit is poor throughout the entire refinement range, but significantly poorer
in the shaded region below 5 \angstrom, as evident from the difference curve. However, this should not come as a surprise if dimers exist in the ITT phase; the \i41amd\ model does not allow for the presence of a nearest neighbor Rh-Rh bond that is significantly shorter than the average bond. The average tetragonal distortion obtained by Le Bail fitting results in a small splitting (less than 0.02 \angstrom) of the nearest-neighbor Rh-Rh distance, but cannot account for the significantly shorter Rh-Rh bond that is associated with dimer formation, which corresponds to a distortion that is nearly an order of magnitude larger.

Fits of the \i41amd structural model to experimental PDFs were carried out throughout the
temperature range studied. Since the dimers cannot be accounted for by this model, their presence will result in an increase in the refinement residual, $R_{w}$. Assuming that the short-bond distortion is the major contributor to the poor fit, $R_{w}$, which is a sensitive measure of the goodness-of-fit, can be used to track this distortion. Figure~\ref{fig:PDFfit} (b) shows $R_{w}(T)$ over the entire temperature range. As expected from the inadequacy of the \i41amd\ model to describe the low temperature state, its value is high in the LTO phase. After an abrupt drop upon crossing into the ITT phase, $R_{w}$ remains unusually large and decreasing with $T$ throughout the ITT phase, consistent with the observations that the local structure is not explained by the average crystallographic model.  While reaching reasonably small values in the HTC phase, $R_{w}(T)$ exhibits a sudden change in slope at $\sim$350~K (Fig.~\ref{fig:PDFfit} (b)), reminiscent of the dimer fraction shown in Fig.~\ref{fig:RDF} (c). Notably, both $R_{w}(T)$ and the dimer fraction (Fig.~\ref{fig:RDF} (c)) resemble $\rho(T)$ (Fig.~\ref{fig:LiRhCharacterization} (a)), suggesting that the transport behavior is marked by the evolution of the dimers. The presence of a dimer bond and its temperature evolution is also seen in the atomic displacement parameter of Rh, and further confirmed through a model independent Gaussian fitting of the PDF data (see Supplementary Material).

These results indicate a complicated and unexpected evolution of the local dimers with temperature.  The observation of the dimer signal in the PDF gives direct evidence for the presence of the CDW along the chains in all the phases, distinct from the sub-band disproportionation that leads to an overall lowering of the cubic symmetry.  This is illustrated in Fig.~\ref{fig:PDFfit}(d) and contrasted with the previous view in Fig.~\ref{fig:PDFfit}(c).  This behavior is different from that observed in \cis\ where the dimers disappear in both the average and local structure at the MIT~\cite{bozin;prl11}.
This view is also supported by a comparison of the specific heat and entropy of \cis\ and \lro. 
The electronic contribution to the entropy is found to be $\sim$2.2~$R$ln2 in \lro, which is significantly larger than 0.03~$R$ln2 in \cis, but comparable to  $R$ln2 in Fe$_3$O$_4$~\cite{shepherd;prb85} (see Supplementary Material). This difference may be explained by electron correlation effects in \lro, which are not present in \cis. 

Additionally, an explanation for the different behavior of the magnetic susceptibility, $\chi$, in \lro\ as compared to \cis\ may reside in our PDF data. In the HTC phase of \lro\ the sample contains few dimers and a balance of itinerant spins; thus it is weakly metallic and $\chi$ is roughly constant with decreasing $T$. At the HTC to ITT phase transition, the spins localize, but many are initially unpaired, resulting in a slight upturn in $\chi$. Then, on cooling through the ITT phase, dimers gradually freeze out. However, significant undimerized Rh$^{4+}$ ions remain (as evidenced by the dimer fraction plotted in Fig.~\ref{fig:RDF}(c)), which contribute to the paramagnetic response and overcome the decrease in $\chi$ due to dimer formation from the other Rh$^{4+}$ ions, thus leading to the observed increase in $\chi$. 
This coexistence of dimerized and undimerized Rh$^{4+}$ ions in \lro\ is consistent with the presence of significant electron correlations leading to frustration in this system.

Our results point to an interesting aspect of the pyrochlore sublattice, namely the corner shared B$_4$ tetrahedra consisting of intersecting
quasi-1D chains of B-ions (Fig.~\ref{fig:model}(a)). This sublattice plays host to a competition between one-dimensional (1D) and three-dimensional (3D) physics.  In a more metallic spinel, such as \cis\, the system can lower its dimensionality through orbital ordering along the chains of B-ions resulting in a 1D chain that is unstable to the formation of a CDW. On the other hand, in a less metallic spinel, such as Fe$_3$O$_4$, the physics is driven by the geometric frustration of ordering different partial charges ($\pm \delta$) on the vertices of the 3D network of tetrahedra~\cite{ander;pr56,senn;nature12,coey;nature04,volja;epl10}. \lro\ lies between these extremes, where the effects of inter-site coulomb interaction and kinetic energy are more comparable, resulting in a competition that leads to the unusual behavior observed in this material.

\section{Acknowledgements}
Sample preparation and characterization work at Argonne was supported by the US DOE, Office of Science, Office of Basic Energy Sciences (DOE-BES) under contract DE-AC02-06CH11357. PDF x-ray experiments, data analysis and modeling were carried out at Brookhaven National Laboratory, supported by DOE-BES under contract DE-AC02-98CH10886.  Use of the National Synchrotron Light Source, Brookhaven National Laboratory, was supported by the DOE-BES under contract No. DE-AC02-98CH10886.

\end{document}


\title{Local Structural Evidence for Strong Electronic Correlations in \lro\ Spinel - Supplementary Material}

\author{K. R. Knox}
\affiliation{Department of Condensed Matter Physics and Materials Science, Brookhaven National Laboratory, Upton, New York 11973, USA}

\author{A. M. M. Abeykoon}
\affiliation{Department of Condensed Matter Physics and Materials Science, Brookhaven National Laboratory, Upton, New York 11973, USA}

\author{H. Zheng}
\affiliation{Materials Science Division, Argonne National Laboratory, Argonne, IL  60439, USA}

\author{W.-G. Yin}
\affiliation{Department of Condensed Matter Physics and Materials Science, Brookhaven National Laboratory, Upton, New York 11973, USA}

\author{A. M. Tsvelik}
\affiliation{Department of Condensed Matter Physics and Materials Science, Brookhaven National Laboratory, Upton, New York 11973, USA}

\author{J. F. Mitchell}
\affiliation{Materials Science Division, Argonne National Laboratory, Argonne, IL  60439, USA}

\author{S. J. L. Billinge}
\affiliation{Department of Condensed Matter Physics and Materials Science, Brookhaven National Laboratory, Upton, New York 11973, USA}
\affiliation{Department of Applied Physics and Applied Mathematics, Columbia University, New York, New York 10027, USA}

\author{E. S. Bozin}
\affiliation{Department of Condensed Matter Physics and Materials Science, Brookhaven National Laboratory, Upton, New York 11973, USA}

\date{\today}

\maketitle

\subsection{\i41amd\ modeling}

As described in the Main Article, fits of the \i41amd\ structural model to \lro\ PDFs were carried out throughout the temperature range studied. Since this model does not allow for short Rh-dimer distances, the presence of dimers is indirectly evidenced through anomolously high values of the refinement residual, $R_{w}$.
%
%
\begin{figure}
\includegraphics[width=85mm]{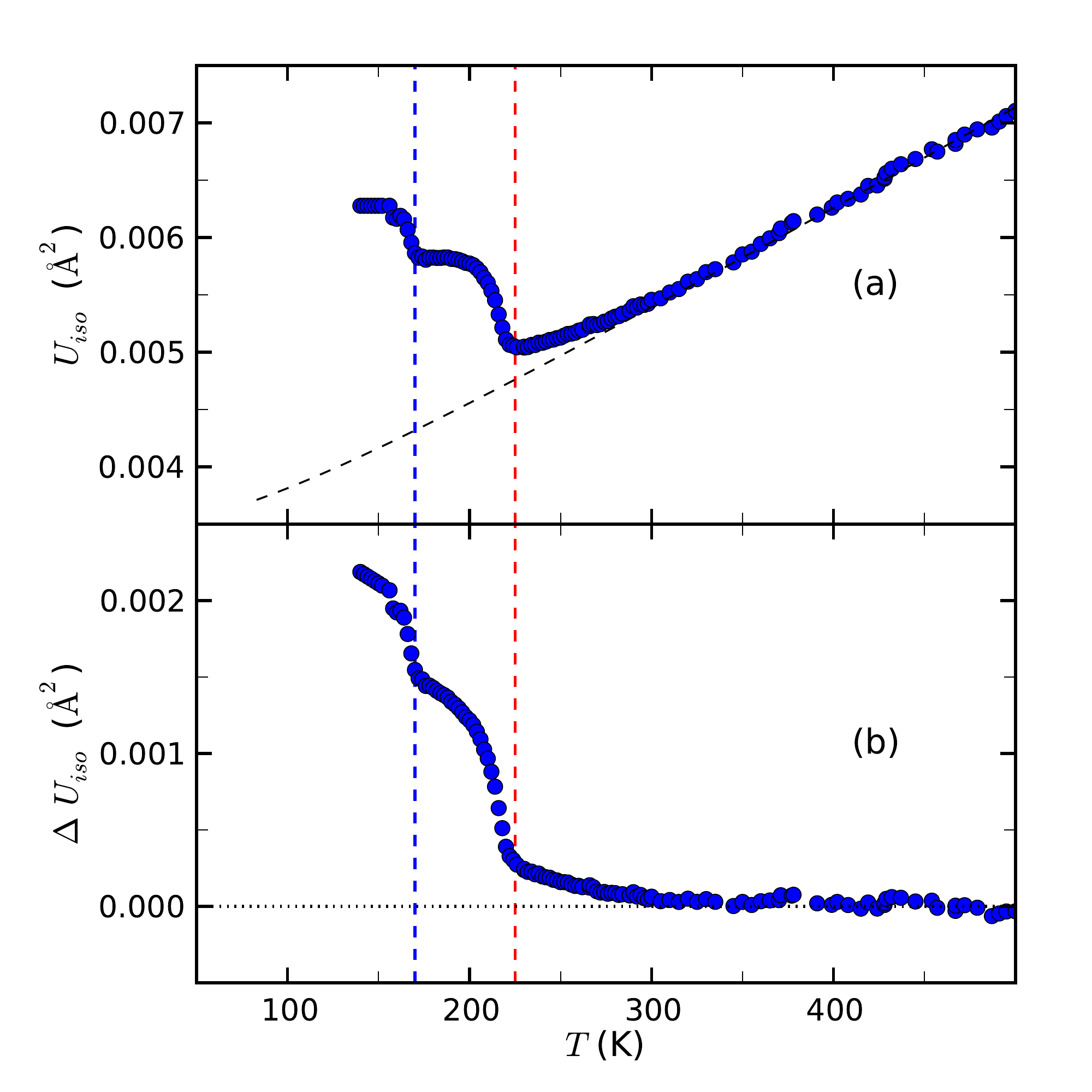}
\caption{\label{fig:RhADP} (Color online)
(a) Temperature dependence of the isotropic atomic displacement parameter (ADP), $U_{iso}$, of Rh obtained from the \i41amd\ model. The black dashed line represents a fit of the Debye model to the high temperature end of the ADP T-dependence. (b) Differential ADP,  $\Delta U_{iso}$, obtained from data shown in (a) by subtracting the Debye curve values from the refined ADPs at all temperatures. Vertical dashed lines mark the phase transitions discussed in the Main Article.}
\end{figure}
%
%
Another sensitive measure of the presence of dimers is the isotropic atomic displacement parameter associated with the Rh crystallographic site. This is shown in Fig.~\ref{fig:RhADP}(a). The canonical temperature dependence of ADPs in materials on cooling 
is their gradual \emph{decrease} with decreasing temperature, reflecting the decrease in thermal-motion. However, this trend is not observed in the Rh-ADP obtained from our structural refinement;
instead, the Rh-ADP is seen to increase with decreasing temperature in the ITT  and LTO phases,
indicating that the \i41amd\ model is inadequate to describe these phases. Additionally, although the refined Rh-ADP decreases on cooling through the HTC phase, it deviates from canonical Debye behavior (see Eqn. 1) between 300 and 350~K.
This behavior reflects the inability of the \i41amd\ model to properly account for the structural features in the data, and the values of Rh-ADP become anomalously high in the model's effort to provide the best fit.

The themperature dependence of ADPs is often described by a Debye-type model as:
 \begin{equation}
U^{2}(T)=\frac { 3h^2} {4\pi^{2}mk_{B}\Theta_{D}}\left(\frac{\Phi(\Theta_{D}/T)}{\Theta_{D}/T} +\frac{1}{4}\right)+U^{2}_{0},
\label{eq:DebyeModel1}
\end{equation}
where
 \begin{equation}
\Phi(x)=\frac{1}{x} \int^{x}_{0} \frac{x'dx'}{e^{x'})-1}.
\label{eq:DebyeModel2}
\end{equation}
Any deviation from this behavior is typically ascribed either to overall inadequacy of the structural model or to the presence of nanoscale features that deviate from the average structure.
Thus, the presence of dimers that are not long range ordered and, hence, unaccounted for by the average structural model, would result in anomolous behavior of the associated ADP.
The dashed black line in Fig.~\ref{fig:RhADP}(a) represents a fit of the Debye-model to the high temperature tail of the Rh $U_{iso} (T)$ curve. As can be seen in the figure, the fit is reasonable at high temperature. However, the data deviate from the canonical behavior described by the model while the sample is still in the HTC phase. In Fig.~\ref{fig:RhADP}(b) we show the difference between the ADP data and the high temperature Debye fit. Again, the characteristic curve shows up, reminiscent of $\rho(T)$, $R_{w}(T)$, and the dimer fraction temperature dependence shown in the Main Article.

\subsection{Gaussian modeling}
%
%
\begin{figure}
\includegraphics[width=85mm]{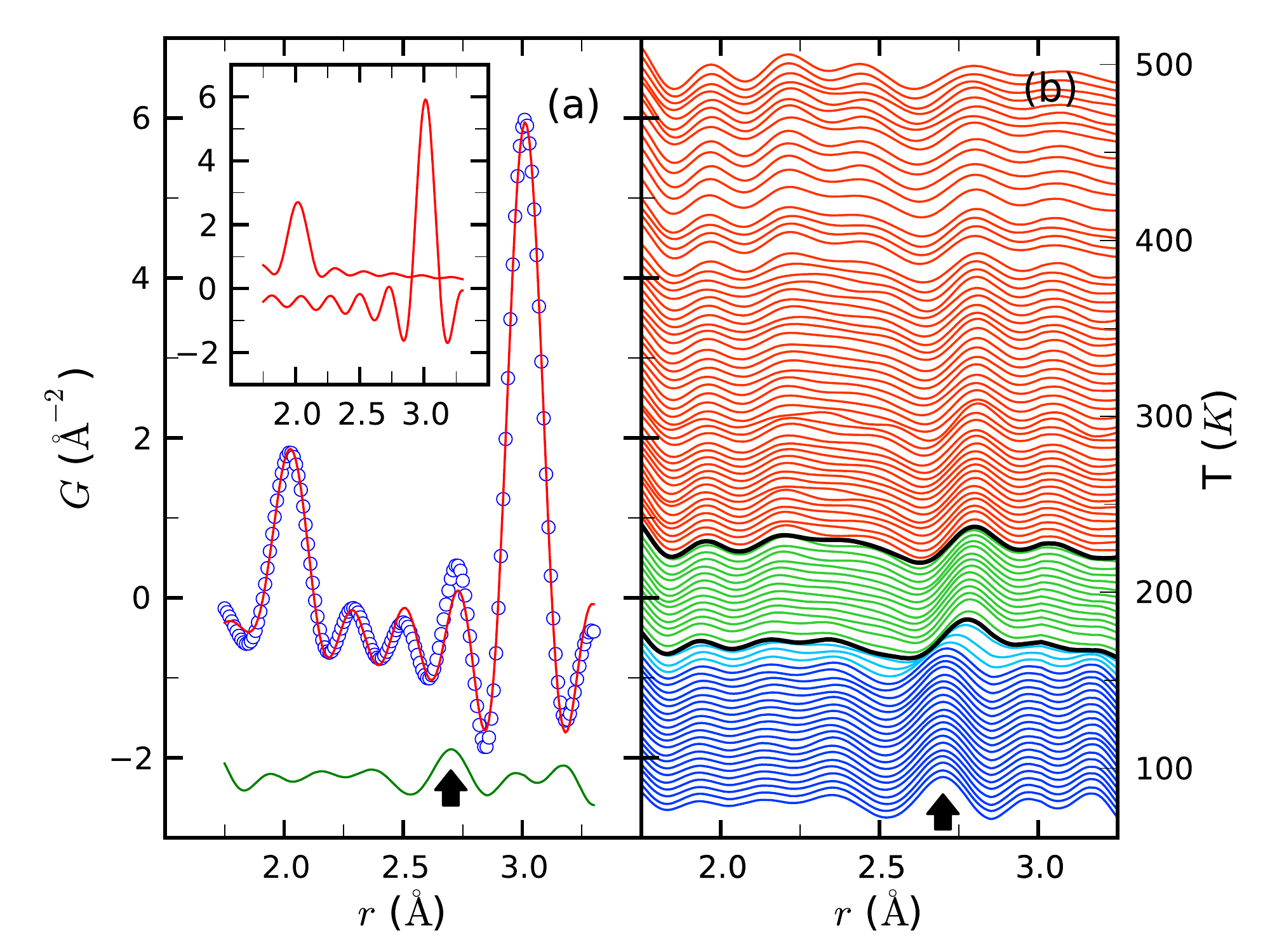}
\caption{\label{fig:2GaussFit} (Color online)
(a) Gaussian fit (red) to 80~K \lro\ PDF (blue circles) in the 1.5-3.5 \angstrom\ range. Difference curve (green) is offset for clarity. Inset shows fit-components. (b) Temperature evolution of the difference curve shown in (a). Arrows in (a) and (b) indicate the short bond peak which is unaccounted for in the 2-Gaussian model. Crystallographic phases are color coded (LTO blue, ITT green, HTC red), while the black profiles mark the transitions.
}
\end{figure}
%
%
To confirm the temperature evolution of the short Rh-dimer bond, a structural model independent assessment of the experimental PDFs was performed; the low-$r$ portion of the \lro\ PDFs were fit with Gaussians convolved with Sinc functions to account for termination effects~\cite{egami;b;utbp13}. The Gaussians were used to fit the range from 1.5~\angstrom\ to 3.5~\angstrom\ to account for the peaks at 2.0~\angstrom\ and 3.0~\angstrom (inset to Fig.~\ref{fig:2GaussFit} (a)). The short bond peak associated with dimerization was intentionally left out of this model, and any unaccounted PDF intensity (such as from the Rh-dimer peak) shows up in the difference curve. In this way, the evolution of the dimer peak may be followed directly by plotting the difference curve obtained by subtracting the best fit of the 2-Gaussian model from the data. A representative fit to the 80~K PDF profile is shown in Fig.~\ref{fig:2GaussFit}(a), with the unaccounted dimer peak in the difference curve marked with arrow.  The temperature evolution of this difference curve is shown in panel (b). As the sample is warmed through the LTO phase, it is clear that the short bond peak is roughly constant in both position and intensity until 160~K at which point it rapidly moves towards higher $r$ in advance of the phase transition at 170~K. The peak then persists through the ITT phase and well into the HTC phase with gradually diminishing intensity.

\subsection{Specific Heat Analysis}
%
%
\begin{figure}
\includegraphics[width=90mm]{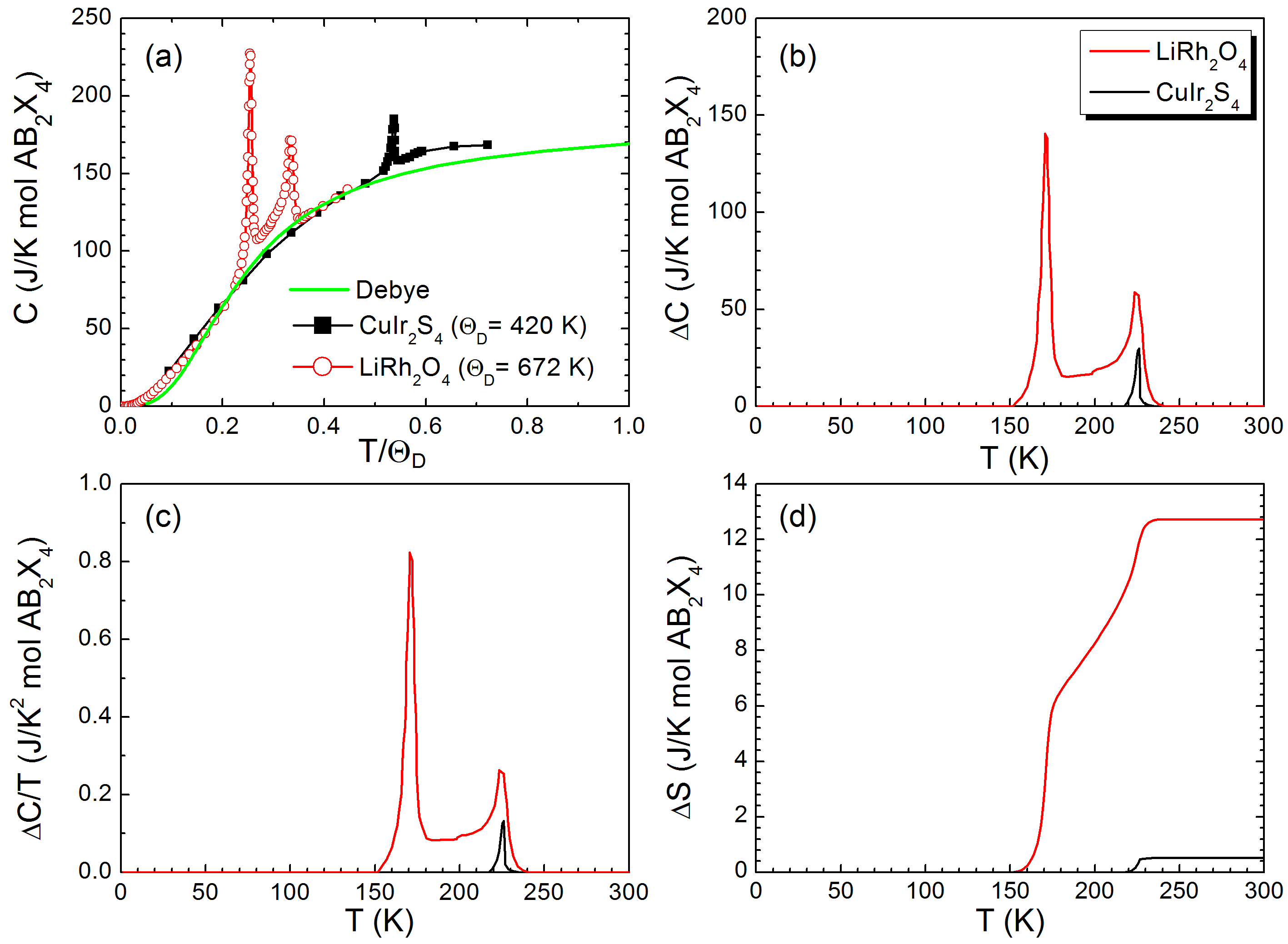}
\caption{\label{fig:Debye} (Color online)
(a) Renormalized-temperature dependence of the specific heat for \lro\ (red circles) and \cis\ (black squares), compared with the Debye model (green line). (b) $\Delta C$, the contribution to the specific heat coming from the phase transitions for \lro\ (red line) and \cis\ (black line). (c) $\Delta C/T$. (d) The corresponding entropy $\Delta S(T)=\int_{0}^{T}\Delta C/TdT$.
}
\end{figure}
%
%
Fig.~\ref{fig:Debye}(a) compares the specific heat, $C$, of \lro\ (obtained from Ref.~\cite{okamo;prl08}) with that of \cis\ (obtained from Ref.~\cite{zhang;jmmm09}). Both are expressed as functions of the renormalized temperature, $T/\Theta_D$, where $\Theta_D$ was obtained by fitting the data over the whole temperature range from 0 to 300~K with the Debye specific heat expression
\begin{equation}
C=9NR\left(\frac{T}{\Theta_D}\right)^3 \int_{0}^{\Theta_D/T}{\frac{x^4 e^x}{(e^x-1)^2}dx}
\end{equation}
The fitted Debye temperatures are $\Theta_{D}$ = 420~K for \cis\ and 672~K for \lro. The larger value of $\Theta _{D}$ for \lro\ can be ascribed to the smaller atomic masses of Li, Rh, and O as compared to Cu, Ir, and S, respectively. 

It is clear that, after scaling, the $C$ curves of \lro\ and \cis\ differ only at the peaks around the transition temperatures. This indicates that the phonon contributions to the specific heat of these materials are rather similar.  Additionally, their values at 300~K are close to $3NR=174.6$~J~mol$^{-1}$~K$^{-1}$ where $N = 7$ (the number of atoms in the chemical formula) and $R = 8.314$~J~mol$^{-1}$~K$^{-1}$ (the universal gas constant), which indicates that phonon contributions dominate $C$ in this temperature regime.  Indeed, it is clear that both curves closely follow the Debye model away from the transition temperatures.

Since $C$ only deviates substantially from the Debye curve near the transition temperaures, it is straightforward to extract the electronic contributions. This was done by subtracting a linear background from the peaks near the transition temperatures as shown in Fig.~\ref{fig:Debye}(b).  Figs.~\ref{fig:Debye}(c) and (d) compare $\Delta C/T$ and the electronic contribution to the entropy, $\Delta S(T)=\int_{0}^{T}\Delta C/TdT$, respectively. Integrating C/T for these peaks shows a significantly larger entropy increase for \lro\ than for \cis.  Since the phonon contributions have already been accounted for, this difference can only arise from contributions to the entropy from the electronic system, including the charge, spin, and orbital degrees of freedom.

It is useful to express the entropy change in the familiar units of $R$ln2, which would be appropriate for a system of localized objects with spin $\frac{1}{2}$ (note that we do not wish to imply by this that \lro\ contains localized spins) and is also the value of the entropy change of the Verwey transition in Fe$_3$O$_4$. The entropy associated with the transition in \cis\ is 0.03 $R$ln2, confirming its weakly correlated nature. However, the corresponding entropy change in \lro\ is about $R$ln2 for the low-T transition and an additional $R$ln2 for the high-T transition. This large electronic entropy indicates the existence of many degenerate states above the transition, a hallmark of strong electron correlation. Although the origin of this large entropy is unclear at this moment (and calls for further theoretical and experimental investigation), it is reasonable to conclude that \lro\ and \cis\ belong to two different regimes in terms of electron correlation.